# Realizing bending waveguides with anisotropic epsilon-near-zero metamaterials

Jie Luo, Ping Xu[b)], Huanyang Chen, Bo Hou, Yun Lai[a)]

*School of Physical Science and Technology, Soochow University, Suzhou 215006, the People's Republic of China*

We study metamaterials with an anisotropic effective permittivity tensor in which one component is near zero. We find that such an anisotropic metamaterial can be used to control wave propagation and construct almost perfect bending waveguides with a high transmission rate (>95%). This interesting effect originates in the power flow redistribution by the surface waves on the input and output interfaces, which smoothly matches with the propagating modes inside the metamaterial waveguide. We also find that waves in such anisotropic epsilon-near-zero materials can be reflected by small-sized perfect magnetic conductor defects. Numerical calculations have been performed to confirm the above effects.

The electromagnetic properties exhibited by metamaterials are remarkable. Indeed, they may provide almost arbitrary effective permeabilies and/or permittivities[1-7]. In previous studies on metamaterials, various interesting phenomena have been discovered, such as negative refraction[1], perfect lens[3], invisibility cloaks[4-7], etc. Metamaterials with near zero parameters are also an important and intriguing class. Recently, epsilon-near-zero (ENZ) metamaterials with permittivity near zero, mu-near-zero (MNZ) metamaterials with permeability near zero, and index-near-zero (INZ) metamaterials with both permittivity and permeability near zero have been extensively studied and various applications have been proposed, such as directive emission devices[8-12], creating subwavelength channels and bends[13-19], tailoring the wave front[20, 21], realizing total transmissions and reflections in a channel by engineering defects[22-25], ultrasensitive defect sensors[25], angular filtering[26], controlling leaky wave radiation[27], etc. One especially interesting application is the bending waveguide effect. For conventional waveguide bends, reflection and distortion of wave front are often dramatic. To minimize the reflection and distortion of waves, many methods have been proposed, such as micro-prism bends[28], photonic crystal bends[29], plasmonic bends[30], bends designed through the principles of transformation optics[31-34], and zero-index materials[13,17-19], etc. Silveirinha *et al.*[13] first proposed using isotropic epsilon-near-zero (IENZ) metamaterials to create subwavelength channels and bends, which was later experimentally realized by Liu *et al.*[17] and Edwards *et al.*[18, 19]. However, in order to ensure high transmission, such isotropic channel is required to be very narrow in width.

In this letter, we investigate the properties of anisotropic epsilon-near-zero (AENZ) metamaterials in which one component of the permittivity tensor is near zero. We find such a material can achieve almost perfect bending waveguides, which does not have the requirement of narrow channel width for high transmission. More interestingly, we can embed a small-sized perfect magnetic conductor (PMC) defect in the AENZ metamaterials to confine the wave propagation. It should be noted that with only one component near zero, such AENZ metamaterials are much easier to achieve than IENZ metamaterials in practice. Possible realizations of such AENZ metamaterials include metal-dielectric multilayered structures[35, 36], metal wire arrays[26, 37], etc.

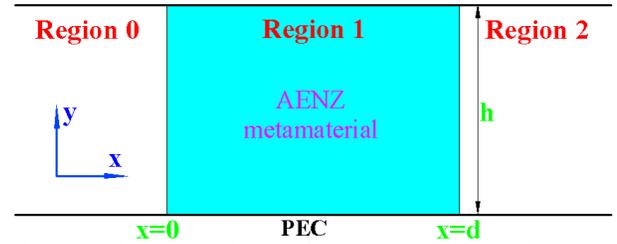

FIG. 1. The schematic graph of a 2D waveguide structure with an AENZ metamaterial (region1). Regions 0 and 2 are filled with air. The parallel lines in the x direction are PEC walls of the waveguide.

Before we study the bending waveguide of AENZ metamaterials, we first investigate the case of a straight waveguide. As illustrated in Fig. 1, we assume a transverse magnetic (TM) polarized wave with magnetic field in the $z$ direction incident from the left port into a two-dimensional (2D) waveguide. Regions 0 and 2 are filled with air and separated by an AENZ metamaterial of a width $h$, a length $d$ in region 1. The AENZ metamaterial has an anisotropic permittivity tensor $\begin{pmatrix} \varepsilon_x & & \\ & \varepsilon_y & \\ & & \varepsilon_z \end{pmatrix}$, and an isotropic permeability $\mu$ being unity. The wavelength of the incident waves in vacuum is $\lambda_0$. The relationship between the electric and magnetic fields in the AENZ metamaterials can be written as,

$$\frac{\partial H_z}{\partial x} = -\varepsilon_y \frac{\partial E_y}{\partial t}, \quad \frac{\partial H_z}{\partial y} = \varepsilon_x \frac{\partial E_x}{\partial t}. \quad (1)$$

Assuming that $\varepsilon_x \approx 0$, from Eq. (1), it is seen that $\partial H_z / \partial y \approx 0$, implying that the magnetic fields are almost constant along the $y$ direction. Thus, the transmission coefficient $t$ can be easily derived from transfer matrix method as[38],

[a)]Electronic mail: laiyun@suda.edu.cn
[b)]Electronic mail: pxu@suda.edu.cn

$$t = \frac{2}{2\cos\varphi - i(a/b + b/a)\sin\varphi}, \quad (2)$$

where $\varphi = 2\pi abd/\lambda_0$ with $a = \sqrt{\mu}$, $b = \sqrt{\varepsilon_y}$. It can be immediately seen that the transmission is an oscillating function of length $d$, and if $\varepsilon_y = \mu$, the reflected waves will vanish due to impedance match with air.

In contrast, for an IENZ metamaterial ($\varepsilon_x = \varepsilon_y \rightarrow 0$) channel of length $d$ and width $h$, the transmission can be written as,

$$t = \frac{2}{2 - i\mu k_0 d}, \quad (3)$$

which implies that the transmission through an IENZ metamaterial channel decays when length d increases. Only when the total area of the channel is very small, the transmission coefficient will tend to unity[13, 17-19]. This represents a big difference between AENZ and IENZ metamaterials.

With the above results for a straight waveguide of AENZ metamaterials, it would be interesting to investigate the cases for bending waveguides. We perform numerical simulations using the finite-element-method (software COMSOL Multiphysics). The structures and results are shown in Fig. 2.

Firstly, suppose that incident waves ($\lambda_0 = 5mm$) impinge onto the AENZ metamaterial channel ($h = 6mm$, $\varepsilon_\theta = 0.001$ and $\varepsilon_r = \mu = 1$) from the left port. $\varepsilon_\theta$ and $\varepsilon_r$ represent the permittivities in the tangential and radial directions for the bent AENZ metamaterial, respectively. The equality of $\varepsilon_r$ and $\mu$ ensures the impendence match between the air and the AENZ slab, leading to zero reflection. Moreover, the near-zero $\varepsilon_\theta$ ensures the uniformity of magnetic fields in the radial direction. As a consequence, the AENZ metamaterial is capable of bending and almost totally transmitting waves in a channel, irrespective of the bending angle. In Fig. 2(a), we show the case of bending the straight AENZ metamaterial channel into a semicircular channel, as illustrated. It is evident that the incident waves can totally transmit through the semicircular AENZ channel. Similarly, nearly perfect bending channels of other shapes, e.g., 60-degree bend, 90-degree bend, and "S" shape, can be made with the AENZ metamaterial.

For comparison, we note that if the AENZ metamaterial is replaced by an IENZ metamaterial ($\varepsilon_\theta = \varepsilon_r = 10^{-3}$), most of the incident waves are blocked and only a very small part can be transmitted through due to the large geometrical area of the channel (see Fig. 2(b)). Figure 3(c) displays the transmission fields in a hollow bending waveguide, showing great distortion in the wave front of transmitted waves. At last, we show that the loss in such metamaterials will not destroy the bending effect. Figure 2(d) presents the field distribution when the waveguide bend is filled with lossy AENZ metamaterials ($\varepsilon_\theta = 0.001 + 0.01i$). It is evident that most of the incident energy can still be transmitted through the bending waveguide and the distortion caused by the absorption is small.

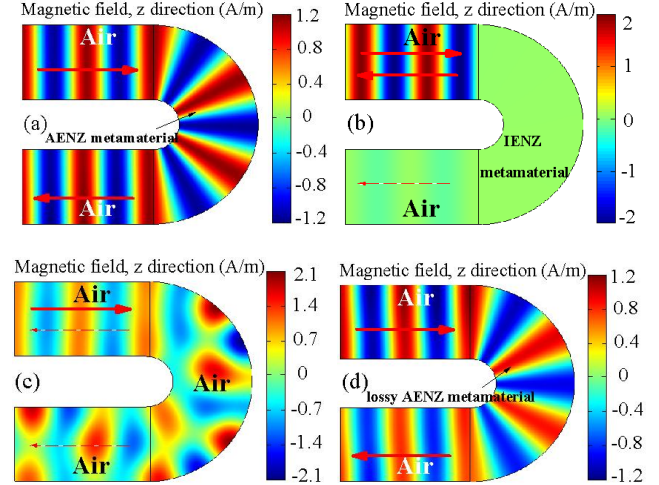

FIG. 2. Magnetic field distribution in a waveguide with the bend part filled with (a) an AENZ metamaterial ($\varepsilon_\theta = 0.001$, $\varepsilon_r = 1$) without defects, (b) an IENZ metamaterial ($\varepsilon_\theta = \varepsilon_r = 0.001$), (c) an AENZ metamaterial with a small PMC defect, and (d) a lossy AENZ metamaterial ($\varepsilon_\theta = 0.001 + 0.01i$, $\varepsilon_r = 1$).

Fig. 2(a) and 2(d) show that the zero index guarantees the wave front to propagate along the bending waveguide. However, to rigorously understand the above interesting results, it is helpful to bend the waveguide into a straight one by transformation optics. According to a coordinate transformation similar to that in Ref. [32], i.e. $x' = R\theta$, $y' = g(r)$, $z' = z$, we obtain the following: $\varepsilon_x = \varepsilon_\theta R/[g'(r)r]$, $\varepsilon_y = \varepsilon_r g'(r)r/R$, $\mu_z = \mu r/[g'(r)R]$. Here we take $g(r) = r$ and $g'(r) = 1$. Considering $\varepsilon_\theta \approx 0$ and $\varepsilon_r = \mu = 1$, we have $\varepsilon_x \approx 0$ and $\varepsilon_y = \mu_z = y'/R$. This means that the homogeneous bending waveguide is equal to a straight waveguide with y-dependent parameters. We should note that even in such a y-dependent waveguide, the impedance $\sqrt{\mu_z/\varepsilon_y} = 1$ is matched everywhere. However, the refractive index $\sqrt{\varepsilon_y \mu_z} = y'/R$ is y-dependent, which usually indicates that the wavelength is also a function of y. However, such a y-dependence also brings longitudinal mode $E_x$ into the eigenstates. The coupled $E_x$ and $E_y$ form a new type of propagating waves inside the metamaterial waveguide with a unified new wavelength irrespective of y, as guaranteed by the translation invariance in the $x$ direction. In Fig. 3, we demonstrate the transmission through such a y-dependent waveguide with $\varepsilon_x = 0.001$ and $\varepsilon_y = \mu_z$ both linearly varying from 0.25(bottom) to 1(top). Fig. 3(a) shows that the magnetic field $H_z$ is almost uniform in the y direction in spite of the y-dependent refractive index. This is due to the near-zero $\varepsilon_x$. In addition, the near-zero $\varepsilon_x$ ensures $\partial(\varepsilon_y E_y)/\partial y \approx 0$, implying that the amplitude of $E_y$ decays almost linearly with y, as shown in Fig. 3(b), and thus the energy flow in the x direction ($E_y \times H_z$) is mostly confined in the low-$\varepsilon_y$ region, as shown in Fig. 3(c). However, for the incident waves, we have almost a constant $E_y$ and a uniformly-distributed energy flow along the y axis. How do they match with each other? To resolve this issue, we plot the $E_x$ map in Fig. 3(d). Interestingly, we find strong surface waves induced by the near zero permittivity on the input and output interfaces of the AENZ metamaterial. Such strong surface waves can redistribute the energy flow on the interfaces to smoothly connect with the propagating modes inside the AENZ metamaterial, and therefore, result in the high transmission coefficient. In Fig. 3(e), we plot the energy flow in the y direction. The existence of giant power flow at the interfaces of the air and the AENZ slab clearly confirms the function of such surface waves. We have tried samples of various length. Our numerical calculations indicate that the transmission rate is always larger than 95%, almost independent of the sample length. In fact, as long as $\varepsilon_y = \mu_z$, the y-dependence of the waveguide also does not change the transmission much.

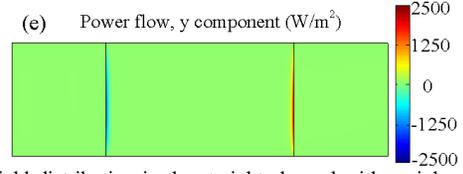

FIG. 3. Field distribution in the straight channel with an inhomogeneous AENZ metamaterial slab characterized by $\varepsilon_x = 0.001$, and $\varepsilon_y = \mu_z$ both linearly varying from 0.25 (bottom) to 1 (top). (a) Magnetic field in the z direction. (b) y components of the electric field. (c) Power flow in the x direction. (d) x components of the electric field. (e) Power flow in the y direction.

Finally, we demonstrate the blocking effects of PMC defects or walls in AENZ metamaterials. In Fig. 4(a), we embed a TM polarized line source in the center of the AENZ metamaterial. The AENZ metamaterial enforces the uniformity of the magnetic field in the y direction. As a result, the wave front of the radiation is changed to be planar[9, 11, 12], as displayed in Fig. 4(a). Interestingly, due to the uniformity of the field, a small PMC defect, which enforces the magnetic field to be zero, is capable of blocking propagation waves, as presented in Fig. 4(b). This is in contrast to the diffraction of fields around a small object in the free space. By changing the PMC defect into PMC wall, the effect also exists, as shown in Fig. 4(c). Therefore, the propagation of electromagnetic waves can be easily controlled by PMC defects in such AENZ metamaterial, even within a confined region (see Fig. 4(d)). In contrast, an IENZ metamaterial with PMC defects will enforce the magnetic field in the whole region to be zero, that is, it cannot support any propagating waves in any direction.

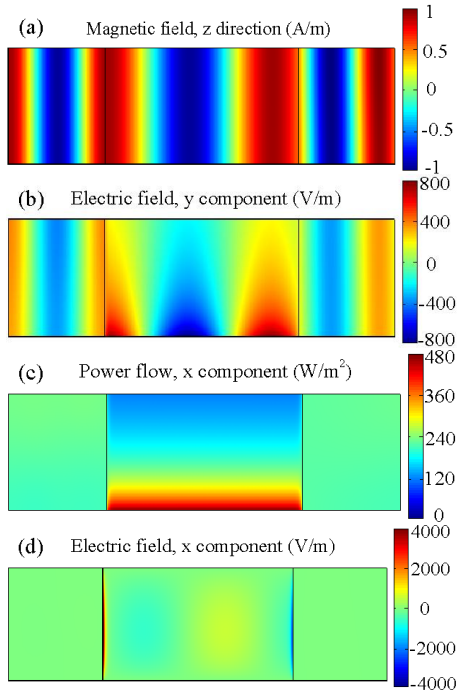

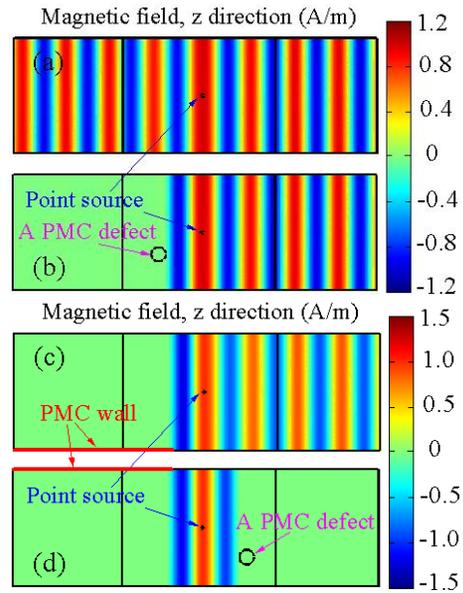

FiG. 4. Magnetic field distribution for a line source along the z direction embedded in the center of an anisotropic ENZ metamaterial. (a) No defect is added. (b) A PMC defect is placed to the left of the line source. A PMC wall is used instead of its original PEC wall in the left without (c) and with (d) a PMC defect embedded to the right of the line source.

In conclusion, we show that AENZ metamaterials in which one component of the permittivity tensor is near zero can be applied to construct nearly perfect bending waveguides and control wave propagation with small-sized PMC defects. One especially interesting finding is that the near-zero-permittivity-induced surfaces waves can redistribute the energy flow properly such that the incident fields match well with the eigenstates inside the AENZ metamaterial waveguide. This mechanism results in high transmission rate in bending waveguides. Since AENZ metamaterials can be easily realized in various metamaterials such as metal-dielectric multilayered structures[35, 36], or metal wire arrays[26, 37], our findings indicate a practical method to achieve perfect bending waveguides and control wave propagation. This work is demonstrated for AENZ metamaterials in the TM polarization, but our theory also works for anisotropic mu-near-zero metamaterials (e.g. Ref. [12]) in the TE polarization.

This work was supported by the State Key Program for Basic Research of China (No. 2012CB921501), National Natural Science Foundation of China (No. 11104196, 11004147, 11104198), Natural Science Foundation of Jiangsu Province (Grant No. BK2011277, BK2010211) and a Project Funded by the Priority Academic Program Development of Jiangsu Higher Education Institutions (PAPD)